\documentclass[aps,prb,twocolumn,showpacs]{revtex4}
\usepackage{graphicx}
\usepackage{dcolumn}
\begin{document}
\title{Structure of Bi nanolines: using tight-binding to search
parameter space}
\author{D.R.Bowler} 
\email[]{david.bowler@ucl.ac.uk}
\affiliation{Department of Physics and Astronomy, University College
London, Gower Street, London, WC1E 6BT U.K.}
\author{J.H.G.Owen}
\email[]{james.owen@aist.go.jp} 
\affiliation{Nanotechnology Research Institute, National Institute of
Advanced Industrial Science and Technology (AIST), AIST Tsukuba
Central 2, 1-1-1 Umezono, Tsukuba, Ibaraki 305-8568, Japan}

\date{\today}

\begin{abstract}
We describe how we have used tight binding calculations as a quick,
efficient tool to search for possible structures of Bi nanolines on
Si(001).  After identifying promising candidate structures, we have
concentrated on these with \textit{ab initio} electronic structure
techniques.  The energetics of the tight binding are shown to be in
good agreement with the density functional calculations and with
experimental observations, and have proved invaluable in the search
for a structure, validating the use of tightbinding as a search tool.
\end{abstract}

\maketitle

\section{\label{sec:Intro}Introduction}

When confronted with a new structure observed in experiment, or when
trying to find possible pathways for reactions or diffusion events,
there is often a need for performing electronic structure calculations
on many different structures rather quickly, in order to search
through parameter space efficiently.  However, the need is not just
for speed, but also for reasonable accuracy, as possible candidates
must not be discarded out of hand.  In this paper, we explore the use
of tight binding as a tool to perform a search for the structure of Bi
nanolines on Si(001), and consider future directions for these types
of search using techniques developed over the last few years.

If a Bi-covered Si(001) surface is annealed at 570-600$^\circ$C, most
of the Bi desorbs, but a small amount self-assembles to form
nanolines, 1.5~nm wide and often over 400~nm
long\cite{Miki1999a,Miki1999b}.  These nanolines have great promise as
templates for nanowires\cite{Owen2001a}, when metals such as Al or Ga
are deposited on them.  However, to understand and control these
processes and the growth of the nanolines, it is important to know
their atomic structure; earlier proposals for the
structure\cite{Miki1999a,Bowler2000a,Naitoh2000} have been shown to be
wrong by recent experimental results, and a valid structure has only
recently been found using a combination of STM and electronic
structure modelling\cite{Owen2001b}.  We describe the search for the
structure, and the tools used for that search, in this paper.

Tight binding\cite{Goringe1997} is a compromise between accuracy (it
makes approximations, but preserves quantum mechanics) and
computational efficiency (simulations of several hundred atoms can be
run on a typical desktop workstation in a few hours).  On
semiconductor surfaces, where a rich variety of reconstructions exist,
depending on a subtle interplay of electronic and geometric effects,
the use of a quantum mechanical technique is vital if correct answers
are to be found.  We have developed parameterisations for Si(001), H
on Si(001)\cite{Bowler1998a} and Bi on Si(001)\cite{Bowler2000a}, and
performed simulations on hydrogen
diffusion\cite{Owen1996,Bowler1998b,Bowler2000b}, gas-source growth of
Si(001)\cite{Owen1997a,Owen1997b,Owen1997c,Bowler1998c,Bowler1998d}
and Bi on Si(001)\cite{Miki1999a,Miki1999b,Bowler2000a,Owen2001b} and
have found that tight binding produces results which are in good
agreement with both experiment and first principles techniques.  In
this paper, we further explore the use of tight binding as a tool to
search through parameter space.

The problem we were faced with was that of finding the structure of
the Bi nanolines.  Various criteria had been found from experiment
which possible candidates needed to fulfil: the lines must have a high
kinking energy; the lines must occupy four Si dimers but lie between
the substrate dimers; they must be more stable than the
Bi-(2$\times$n) high coverage phase; and they must repel missing dimer
defects and steps down.  We needed an accurate, quantum mechanical
technique that would allow us to test many possible structures
quickly, allow us to have confidence in new reconstructions and
distorted bonds, and that would also allow us to model large unit
cells (to test kinking and defect repulsion) -- a linear scaling or
${\mathcal O}(N)$ tight binding technique was found to be perfect.
The rest of the paper describes the search: we start by describing the
technical details of the modelling; we then consider the problem of
modelling Bi on Si(001) and comparing structures with different
amounts of Bi, and look at the stability of the Bi-(2$\times$n) high
coverage phase; we present the details of various more or less
successful models, and compare their energies to first principles
calculations, and then we conclude.

\section{\label{sec:tech}Technical Details}
We performed tight binding simulations using a linear scaling
technique (where the computational effort scales linearly with the
number of atoms, not with its cube as is true for traditional
techniques\cite{Goedecker1999}) as large cells, of up to several
thousand atoms, were necessary to test the kinking and dimer repulsion
properties of the nanoline structures.  Specifically, we used an
implementation of the Density Matrix Method (DMM)\cite{Li1993} by
Goringe\cite{Goringe1995}.  We used tight binding parameterisations
for Si-Si bonds developed specifically for Si(001)\cite{Bowler1998a}
and for Bi on Si(001)\cite{Bowler2000a}.  The unit cells were always
ten layers deep, with the bottom two layers constrained to lie in
bulk-like positions, and terminated in hydrogen.  For simple structure
calculations, each cell was one dimer row wide (using the p($2\times
2$) reconstruction) and sixteen dimers long (i.e. 384 atoms).  When
calculating kinking energies we used the same size cell, but four
dimer rows wide (i.e. 1536 atoms).  These methods have been used
before with great success for calculations of many thousands of atoms
(e.g. for calculation of step kinking energies on
Si(001)\cite{Bowler1998c}).

For an accurate assessment of promising candidates, we performed first
principles density functional (DFT) simulations within the local
density approximation (LDA) using the VASP code\cite{Kresse1996}.  We
used ultrasoft pseudopotentials, a plane wave cutoff of 150~eV (which
gave energy difference and force convergence) and a
Monkhorst-Pack\cite{Monkhorst1976} grid of $4\times 4\times 1$
\textbf{k}-points. Each unit cell was the same depth and width as the
tight binding cells, but only eight dimers long, again with the bottom
two layers held fixed in bulk-like positions and terminated in
hydrogen.  The change in energy caused by the shorter cell was found
to be less than 0.1~eV; the longer tight binding cells were used
to investigate nanoline-dimer repulsion, which could not be done using
LDA.

A key question to consider when modelling the stability of structures
with different chemical elements is how to compare structures with
different stoichiometries.  As the experiments we are considering were
performed at high temperature on a Si(001) substrate, we can use the
energy of bulk silicon to correct for different numbers of silicon
atoms (though this raises questions about \textbf{k}-point sampling
and error cancellation for the LDA calculations that will be
considered below).  For bismuth, however, we are faced with a harder
problem.  In the next section, we detail the approach used for
calculating the periodicity of the high coverage Bi-($2\times$n)
phase, and extract a sensible reference energy for Bi dimers on
Si(001).

\section{\label{sec:energetics}Energetics of Bi on Si(001)}
At high coverages, Bi forms rows of dimers on Si(001), relieving the
strain due to the large difference in lattice constants by having
missing dimer defects at regular intervals, forming a high coverage,
Bi-(2$\times$n) reconstruction.  As we see Bi nanolines forming within
this (2$\times$n) surface (and we also see that the nanolines are more
stable than it -- they form while it is desorbing) we start by
describing it.  We approach this problem, which will involve different
amounts of Bi, and hence different stoichiometries, in the same way
that we have approached modelling of the Ge/Si(001)
surface\cite{Oviedo2002}.  It is perfectly easy to calculate the
energy of formation of trenches of missing dimers simply by removing
Bi dimers, but to make contact with experimental results we need to
find an energy for these Bi dimers that we have removed.  We shall
assume that the Bi-covered surface is in thermal equilibrium (which is
easy to arrange experimentally), so that for a given number of Bi
atoms we can write the probability of any given arrangement of these
atoms as proportional to $\textrm{exp}\left(E_a/kT\right)$, where
$E_a$ is the energy of the arrangement.  The assumption of thermal
equilibrium greatly simplifies the system that we are considering,
while being perfectly legitimate at a number of different
temperatures.

Now, as we are in thermal equilibrium, we only need know the
\textit{relative} energetics of different arrangements of the given
number of Bi atoms.  It will be useful to define a reference
arrangement which we take to be a non-defective, lattice-matched layer
of Bi covering a certain area of the Si(001) surface (with a boundary
which we assume to be rectangular, but whose shape is arbitrary).  Now
we consider arrangements where we form regularly spaced trenches of
missing dimers; conceptually we start by adding a number of dimers to
the reference system equal to the number that we will remove and then
forming the missing dimer trenches and relaxing.  This conceptual
approach allows us to compare different spacings by using the energy
of Bi dimers in the perfect monolayer to correct for different amounts
of Bi, which we find by calculating the difference in energy between a
clean Si(001) surface and the same surface covered by a monolayer of
Bi.  Using our tight binding parameterisation we find this number to
be -11.41~eV per Bi dimer.

\begin{figure}
\includegraphics[width=\columnwidth]{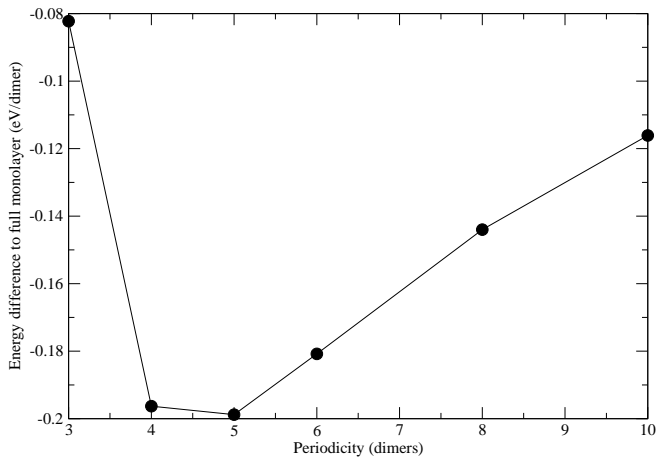}
\caption{Energy difference per dimer between reconstructed surface and
full monolayer for Bi on Si(001) in eV, plotted against periodicity of
reconstruction (where number of Bi dimers is one less than periodicity).
\label{fig:2xn}}
\end{figure}

We show the difference in energy per Bi dimer between a full, strained
monolayer and the reconstructed surface for various values of period,
n, in Fig.~\ref{fig:2xn}.  There is a well defined minimum for periods
of four and five (i.e. three and four Bi dimers) which is in excellent
agreement with experimental observations\cite{Miki1999a,Park1993}.
The modelling of this reconstruction gives us further confidence in
our parameterisation, and presents a limit for the energetics of the
Bi nanoline (which must be more stable than the Bi-($2\times$n)
surface).  Our tight binding simulations find that the energy per Bi
dimer (adsorption energy and surface energy) in this reconstruction is
-11.6~eV at n$=4$ and n$=5$.

When considering structures with different amounts of Bi, we can use
the fact that the nanolines are observed to form in the high coverage
Bi-(2$\times$n) phase to extract a sensible reference energy for Bi
dimers.  We have chosen the energy to be -11.4~eV per dimer, the
energy of the reference phase, and used this where necessary.  As
mentioned above, we use the bulk Si energy to correct for differing
numbers of Si atoms.

\section{\label{sec:models}Details of Successful Models}
In this section, we present details of various of the models
considered, and discuss how well tight binding has performed as a tool
to discriminate between these models in the light of subsequent DFT
modelling.  First, we start by summarising the requirements placed on
the models by experimental observations.

\subsection{\label{sec:expt}Experimental Constraints}
We have been able to distil a number of requirements for the nanoline
structure from our experiments\cite{Miki1999a,Miki1999b,Owen2001a,Owen2001b}:

\begin{itemize}
\item Stability: The lines form at a temperature where the high
coverage (2$\times$n) Bi islands are evaporating, which implies that
the energy per Bi dimer must be larger than these islands (in tight
binding, more than -11.6~eV/dimer as given above).  In RHEED
experiments, the difference in desorption barrier for Bi from the
islands and from the nanolines was determined to be
0.25~eV\cite{Miki1999a}.

\item Width, Registration and Appearance: The nanolines run
perpendicular to the Si(001) surface dimer rows, occupying the space
of four Si dimers (15.36~\AA).  They consist of two dimer-like features
separated by $\sim$6.3~\AA which lie between the Si(001) dimers, but
are in line with the dimer rows.

\item Straightness: Over many hundred observations of many lines (for
instance in previous papers\cite{Miki1999a,Miki1999b}), a kink has
never been seen.  Given the high temperature at which these nanolines
form, it seems likely that a large thermodynamic kink energy is
required, though a kinetic barrier cannot be ruled out.

\item Defect Exclusion Zone: The nanolines are
observed\cite{Owen2001b} to repel defects and down step edges to a
distance of 3-4~nm.  As the defects and steps have a tensile stress
field, we expect that the nanoline model should also have a tensile
stress field.
\end{itemize}

\subsection{\label{sec:theo}Energetics from tight binding and DFT}
Over the course of approximately a year, we have considered many 
different models.  In general, we have first calculated the energy and
kinking energy of possible models using tight binding.  Only if these 
looked promising were they modelled with the more expensive DFT.

\begin{figure}
\begin{center}
\vbox{
\includegraphics[width=0.95\columnwidth,clip]{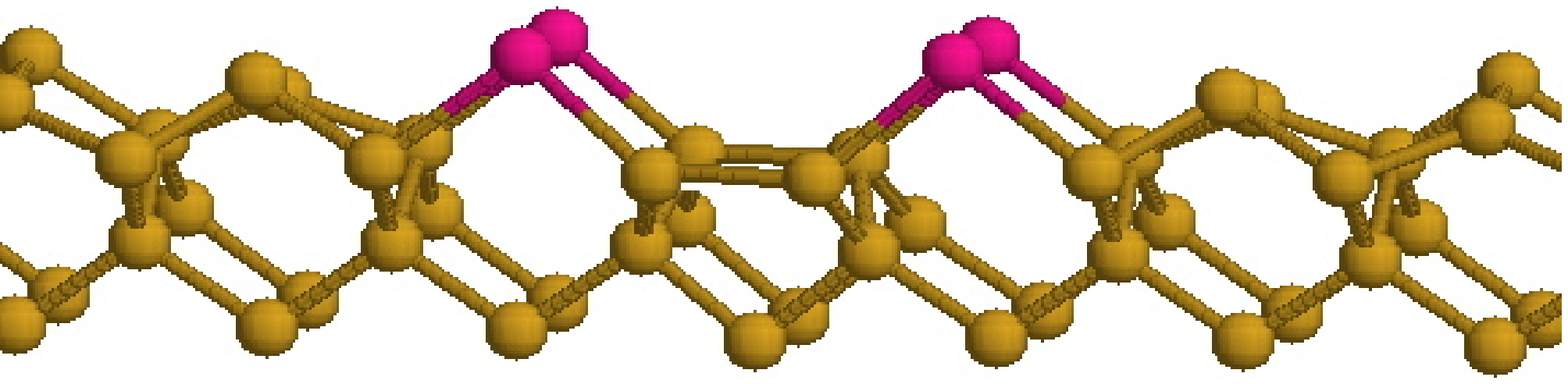}
\includegraphics[width=0.95\columnwidth,clip]{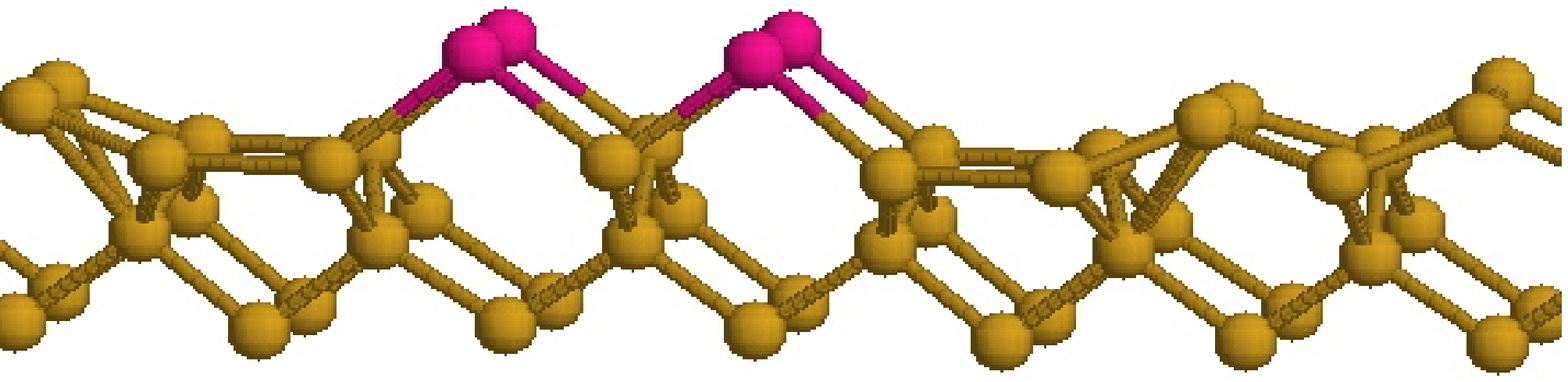}
\includegraphics[width=0.95\columnwidth,clip]{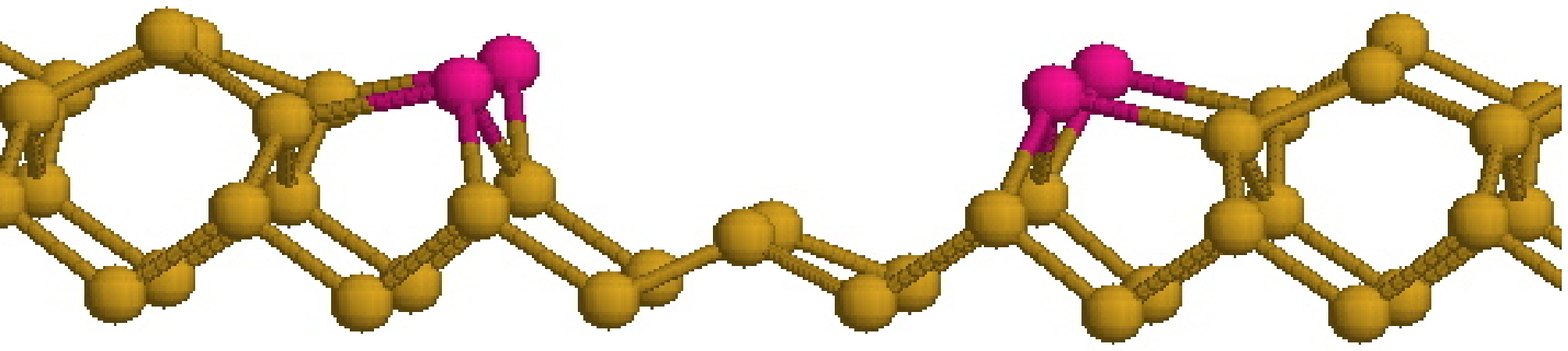}
\includegraphics[width=0.95\columnwidth,clip]{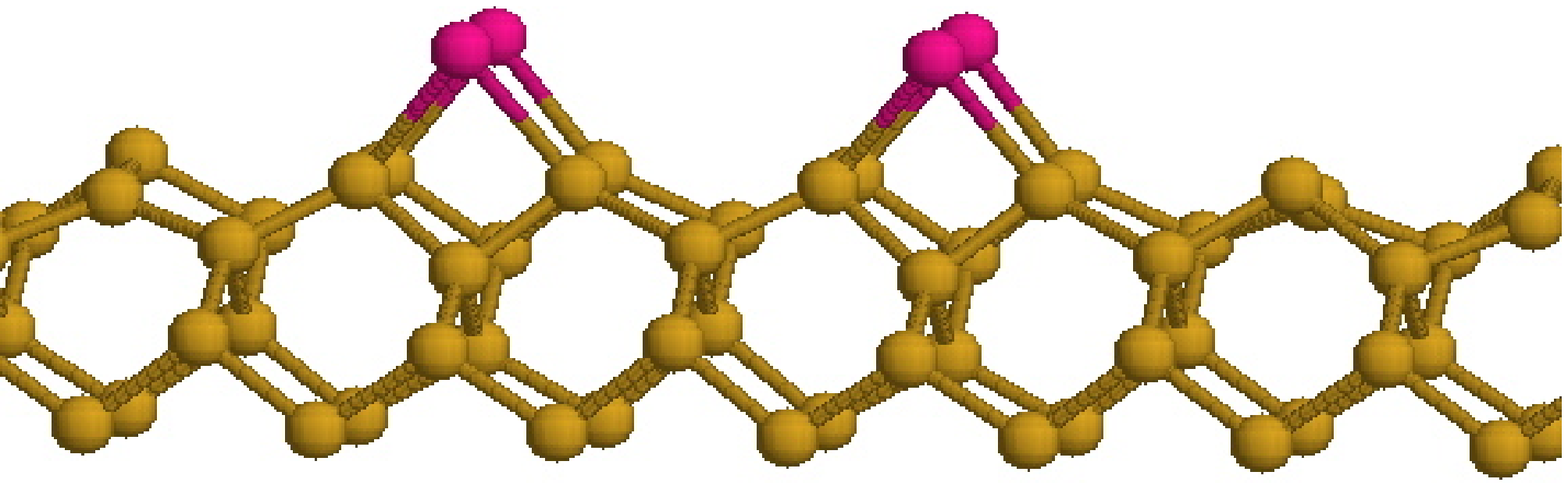}
\includegraphics[width=0.95\columnwidth,clip]{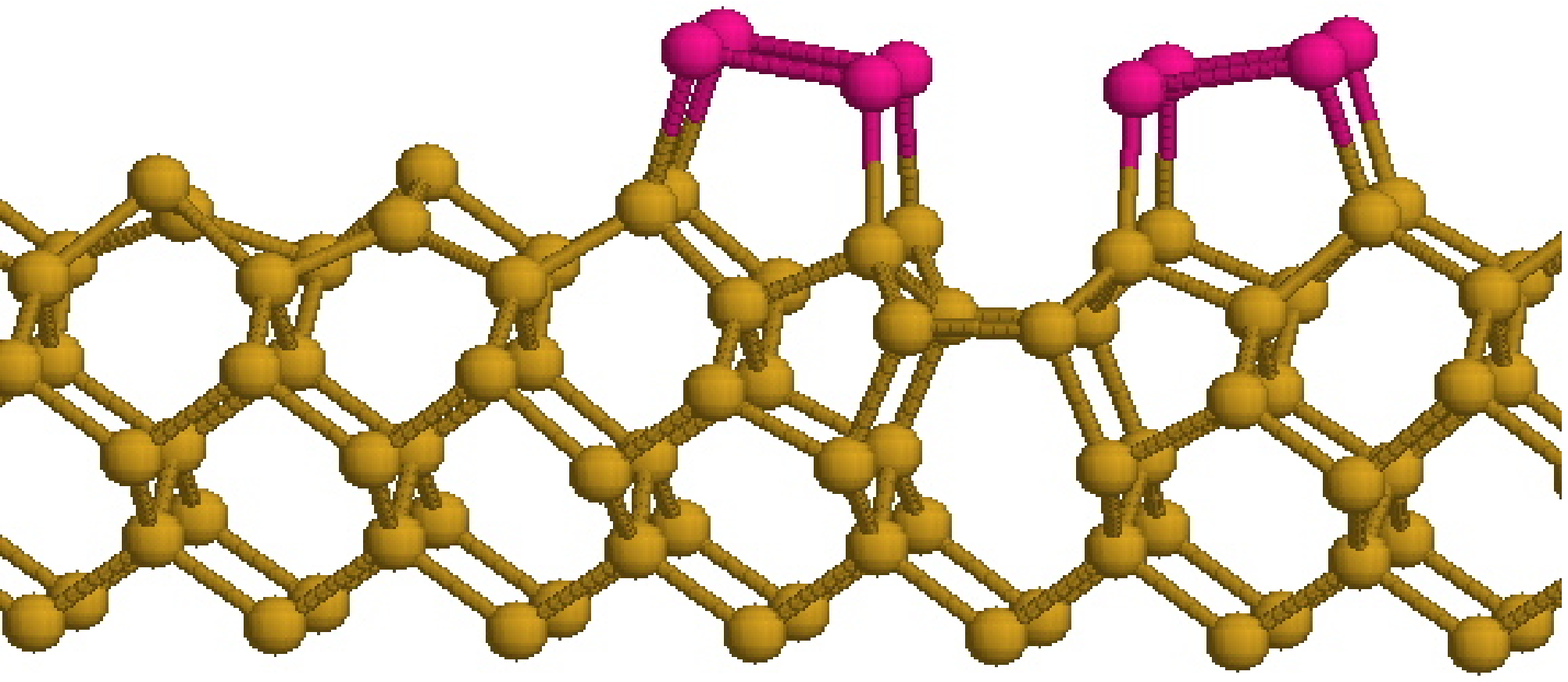}
}
\end{center}
\caption{\label{fig:models}Ball-and-stick models of various
candidate models for the Bi nanoline.  Labels, from top to bottom:
original\protect\cite{Miki1999a}, Two1DVs\protect\cite{Naitoh2000},
B-type 4DV, two ad-dimers and two squares.  Energetics are given in
table~\protect\ref{tab:energy}, and the structures are discussed more
fully in the text.}
\end{figure}

Within this section we present details of five unsuccessful models and
what we believe to be the correct model\cite{Owen2001b}, giving
energies found with both tight binding and DFT.  These results show
that the tight binding calculations certainly ordered the structures
correctly in terms of energy compared to DFT calculations (to within
0.1~eV) even though the absolute energies were sometimes rather too
large, and thus confirm the use of this method as extremely valuable
when searching a large configuration space.  It is also worth noting
that the tight binding simulations were computationally cheap (running
in a matter of hours on a desktop workstation) and also that they
deepened the collaboration between experiment and theory (many tight
binding simulations were run by an experimentalist (JHGO), enabling
detailed discussions of the remaining promising candidates).  As an
illustration, we found that tight binding relaxation calculations on a
large unit cell used for kinking calculation (1536 atoms) took 2 hours
and 4 minutes on a PowerMac G4/867MHz (using open-source g77 and ATLAS
BLAS routines optimised for AltiVec) and 1 hour and 45 minutes on a
Pentium 3/800MHz (using the Intel Fortran compiler for Linux and
either ATLAS BLAS routines or the Intel Math Kernel library).

\begin{figure}
\begin{center}
\includegraphics[width=0.95\columnwidth,clip]{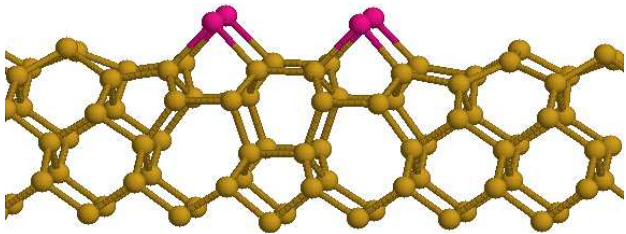}
\caption{\label{fig:core}Ball-and-stick model for the successful
model\protect\cite{Owen2001b}.  There has been significant
reconstruction below the Bi ad-dimers, yielding 5- and 7-membered
rings and changes down to the fifth layer.  However, this structure is
extremely stable in both TB and DFT, and satisfies all experimental
constraints.}
\end{center}
\end{figure}

In Fig.~\ref{fig:models} we show the ball-and-stick models of five
unsuccessful candidates, and in Fig.~\ref{fig:core} we show the final
candidate\cite{Owen2001b} (shown in off-axis views as the structures
are rather complex).  The five unsuccessful models shown are: the
original structure proposed\cite{Miki1999a,Miki1999b,Bowler2000a},
which fitted original STM data but is only 3 Si dimers wide and has a
very low kinking energy (labelled ``Original''); the model proposed by
Naitoh {\it et al.}\cite{Naitoh2000}, which has the wrong spacing
between features, is energetically poor and has a very low kinking
energy (labelled ``Two1DVs''); a structure made from back-to-back
rebonded B-type steps, with Bi substituted in the second layer, which
has the wrong appearance and is not energetically good enough
(labelled ``Btype4DV''); two Bi ad-dimers, which have no kinking
energy and are energetically poor, and have no incentive to remain in
this configuration (labelled ``TwoAdDimers''); and two squares of Bi
on top of a reconstruction similar to the final
structure\cite{Owen2001b} and to the structure of steps on As-covered
Ge(001)\cite{Zhang2001}.

\begin{table}
\caption{\label{tab:energy}Energies of various candidate structures as 
found by tight binding and DFT.  Energies are given relative to the 
three dimer model originally proposed\protect\cite{Miki1999a,Bowler2000a}, 
but now known to be inconsistent with experimental observations.
Each model contains two Bi dimers, and has been corrected for different
numbers of Si atoms as described in the text, except for TwoSquares, 
where the energy is given per square (i.e. equivalent to the others).
Labels are explained in the text, and models are shown in 
Figs.~\protect\ref{fig:models} and \protect\ref{fig:core}.}
\begin{ruledtabular}
\begin{tabular}{lll}
Model       &  TB (eV) & DFT (eV) \\
\hline
TwoSquares  &   1.179  &  1.363 \\
TwoAdDimers &   1.565  &  0.848 \\
Two1DVs     &   1.232  &  0.759 \\
Btype4DV    &   0.611  &  0.564 \\
Btype3DV    &   0.473  &  0.279 \\
Original    &   0.000  &  0.000 \\
DoubleCore  &  -1.904  & -0.720 \\
\end{tabular}
\end{ruledtabular}
\end{table}

We give the energies for these models, along with a variation on the
Btype4DV structure (the same structure but with only a 3 dimer width,
which is energetically rather better, but has the wrong spacing and a
low kinking energy, labelled ``Btype3DV'' and not illustrated) and the
final candidate (labelled ``DoubleCore'') in Table~\ref{tab:energy}.
Where the number of silicon atoms has differed from the clean surface,
we have corrected using an appropriate value for the bulk energy for
silicon.  In the DFT calculations, it is questionable to use simply an
energy calculated from a bulk Si cell, as the \textbf{k}-point
sampling will differ unless fully converged in both cells; we also
expect error cancellation to be better in the same sized unit cells.
We approached the problem from two directions: first, we calculated
the total energy of a clean Si(001) surface using both an eight layer
and a ten layer slab, and used the energy difference to give us a
value for the energy of bulk Si.  We compared this to calculations of
a bulk Si cell which were fully converged with respect to
\textbf{k}-point sampling (we used a $4~\times~4~\times~4$ mesh, and
checked with an $8~\times~8~\times~8$ mesh), and found that the energy
difference less than 0.05~eV per atom.  The energy that we used was
5.96~eV/atom.

From these energies, we can see that the tight binding calculations
have correctly ordered almost all of the structures, though the
energies are sometimes too large.  It is extremely pleasing that it
has performed so well for such a large variety of structures,
especially the final candidate which contains a large stress and
significant reconstruction.  In particular, it can be seen that there
are a wide range of bond lengths (Bi-Bi are generally about 2.9~\AA,
but Bi-Si vary from 2.60 to 2.82~\AA) and angles (from $84^\circ$ to
$106^\circ$) which the tight binding calculations have modelled
correctly.  Bi seems to favour $p^3$ bonding, with 90$^\circ$ bond
angles and a pair of electrons in an s orbital -- and the tight
binding calculations have modelled this well.  The only point where
the ordering is wrong is for the TwoSquares structure, which tight
binding found to be more stable than it should have been.  The Bi-Bi
bonds are generally shorter than found in DFT calculations
($\sim$3.0-3.1~\AA), suggesting that the Bi-Bi bond in TB might be too
strong, accounting for the increased stability of this structure.  The
magnitudes of the TB energies are too large, probably because the
model was fitted to bulk Bi-Bi and Bi-Si bonds, and this environment
is at the surface, allowing Bi to form bonds with angles close to
$90^\circ$ (its preferred bonding angle).

The unsuccessful models were rejected for a number of reasons: they
were energetically poor (Two1DVs, TwoAdDimers, TwoSquares); they had
the wrong appearance (Btype4DV and Btype3DV); they had a low kinking
energy (Original, Two1DVs, TwoAdDimers, Btype3DV); they had the wrong
spacing of Bi dimers (Two1DVs, Btype3DV, Btype4DV); they had the wrong
registry of Bi dimers relative to the substrate (Original, Two1DVs);
and they were less stable than the high coverage Bi-(2$\times$n)
surface (Two1DVs, TwoAdDimers, Btype3DV, Btype4DV, TwoSquares).

The final candidate (DoubleCore) has the correct appearance, a high
kinking energy (3.75~eV from tight binding calculations), the correct
spacing of features (6.3~\AA), a high repulsion for defects and down
step edges (out to 6 or 8 dimers or $\sim$3~nm), and is more stable than
the high coverage Bi-($2\times$n) surface.  We have discussed this
structure in more detail elsewhere\cite{Owen2001b}.

\section{\label{sec:conclusions}Discussion and Conclusions}
The key drawback with tight binding is that a parameterisation must be
created for every interaction, and this is frequently a time-consuming
procedure.  Another extremely important problem for tight binding is
that of \textit{transferrability}: how well a parameterisation
describes interactions that are dissimilar to the environment used for
fitting.  The Bi-Bi and Bi-Si parameterisation used was created simply
and quickly\cite{Bowler2000a}, using the universal tight binding
parameters of Harrison for Bi-Bi hopping\cite{Harrison1980} and
fitting the scaling to the bulk modulus only; the parameterisations
have only been used in environments close to those used for fitting.
This simple procedure (which has also been applied to Ge-Ge and
Ge-Si\cite{Bowler2002}) has been shown here to be remarkably
effective.

However, there is clearly a need for more reliable ways to explore
parameter space in situations such as that encountered in this paper,
and developments in the field known as \textit{ab initio} tight
binding are providing the answers.  Starting with the Harris-Foulkes
theory\cite{Harris1985,Foulkes1989}, the link between DFT methods
using local, atomic-like orbitals and tight binding was shown.  This
was quantified in an elegant demonstration that tight binding could be
derived from DFT via a well-defined series of
approximations\cite{Sutton1988}.  Ideas such as these have led to
different approaches to improving and generalising tight binding.

Typically, confined atomic orbitals (found by solving the
Schr\"odinger equation within a confining potential)\cite{Sankey1989}
are used as a basis set, adding orbitals beyond the minimal set as
necessary for completeness.  The problem with any approach like this
is to decide where to make approximations.  There are two key strands
to these approximations: to neglect self consistency (but to retain
three-centre integrals)\cite{Sankey1989}; and to add self-consistency
but to neglect three-centre terms\cite{Porezag1995}.  Of course, there
have been developments such as adding self-consistency to the
three-centre approach\cite{Demkov1995}.

The ultimate generalisation of these ideas lies in local orbital-based
DFT methods such as \textsc{Siesta}\cite{Sanchez1997},
\textsc{Plato}\cite{Horsfield1997,Kenny2000} and
\textsc{Conquest}\cite{Bowler2000c}, which point the way forward, as
they enable the creation of the matrix elements (equivalent to a
parameterisation) accurately, both in advance and in real time: the
link between tight binding methods and DFT-based methods is becoming
ever stronger.  Clearly the level of approximation chosen needs to
trade off against the size and complexity of the system being
modelled.

In conclusion, we have presented tight binding and DFT calculations
for the high coverage Bi-($2\times$n) surface which are in good
agreement with experiment, and we have explored a variety of possible
structures for Bi nanolines.  Tight binding calculations have been
shown to agree well with DFT calculations in terms of ordering
structures, and enabled us to perform quick, efficient searches of
parameter space.  In particular, it has enabled us to find an
extraordinary structure for the Bi nanoline, which in turn sheds light
on the role of surface stress in self-assembly of nanostructures on
semiconductor surfaces.

\begin{acknowledgments}
DRB thanks the UK Engineering and Physical Sciences Research Council
and the Royal Society for support through fellowships.  JHGO is
supported by the Japanese Science and Technology Agency (JST) as an
STA Fellow.  This study was performed through Special Coordination
Funds of the Ministry of Education, Culture, Sports, Science and
Technology of the Japanese Government (Research Project on active
atom-wire interconnects).  We are happy to acknowledge useful
discussions with Dr. K.Miki, Prof. G.A.D.Briggs, Dr. W.McMahon and
Prof. M.J.Gillan.  DFT calculations were carried out in the HiPerSPACE
Centre at UCL.
\end{acknowledgments}

\bibliography{TBSearch}
\end{document}